\documentclass[a4paper,twoside]{article}
\newcommand{\affil}[1]{$^{\rm #1}$}
\usepackage[authoryear]{natbib}
\bibpunct{(}{)}{;}{a}{}{,}
\usepackage{graphicx}
\date{} 
\title{\large\bf\flushleft Data storage, processing and visualisation for the ATCA}
\author{\parbox{\textwidth}{\flushleft
\vspace{-0.5cm}
{\it T. Murphy\affil{A,C,D}, P. Lamb\affil{B}, C. Owen\affil{C}, and M. Marquarding\affil{C}}\\
\vspace{0.4cm}
{\small \affil{A}\,School of Physics, University of Sydney, NSW, 2006, Australia}\\
{\small \affil{B}\,CSIRO ICT Centre, Canberra, ACT, 2601, Australia}\\
{\small \affil{C}\,Australia Telescope National Facility, Epping, NSW, 1710, Australia}\\
{\small \affil{D}\,Email: tara@physics.usyd.edu.au}}}
\begin{document}
\twocolumn[
\begin{changemargin}{.8cm}{.5cm}
\begin{minipage}{.9\textwidth}
\vspace{-1cm}
\maketitle
%
%
\small{\bf Abstract:} We present three Virtual Observatory tools developed at the ATNF for
the storage, processing and visualisation of ATCA data. These are the Australia 
Telescope Online Archive, a prototype data reduction pipeline, and the Remote Visualisation
System. These tools were developed in the context of the Virtual Observatory and were 
intended to be both useful for astronomers and technology demonstrators.
We discuss the design and implementation of these tools, as well as issues that should be 
considered when developing similar systems for future telescopes.


\medskip
\medskip
\end{minipage}
\end{changemargin}
]
\small

\section{Introduction}
The so-called data explosion in astronomy promises exciting new scientific developments, 
but brings with it many technical challenges, in collecting, storing, transporting,
processing and visualising data. 
Virtual Observatories (VO) have developed to meet some of these technical challenges.
Falling under the broad area of e-science (which incorporates other scientific domains
facing similar challenges, such as genetics and particle physics) the aim of Virtual 
Observatory research is to provide the tools necessary for dealing with this data.

The Australian Virtual Observatory (Aus-VO)\footnote{{\tt http://www.aus-vo.org}} 
was started in 2003 with the aim of both contributing to the international VO effort, 
and developing tools of use to Australian astronomers. 
Australia has many areas of expertise (for example radio astronomy) and it makes sense
to focus our efforts on providing modern tools for working in these areas.
In this context the ATNF decided to develop a range of tools for storing, processing 
and visualising data from the Australia Telescope Compact Array. 
The aim was that these tools would be useful to astronomers now, and at the same time
let us explore the technology that would be necessary for developing software for 
future telescopes such as the Square Kilometre Array (SKA). 

This paper is based on a talk given at the ASA Annual Meeting in Sydney in July 2005.
After giving some background about the ATCA and the Virtual Observatory, we discuss three 
tools developed at the ATNF over the last few years.
Firstly the Australia Telescope Online Archive which contains all of the data collected
so far by the ATCA; secondly a prototype data reduction pipeline for ATCA data; and 
finally the Remote Visualisation System for viewing large datasets.

\section{The ATCA}
The Australia Telescope Compact Array (ATCA)  is an east-west earth-rotation synthesis 
interferometer, with six 22 m antennas on a 6 km baseline. 
It has been in operation at Narrabri since 1990. 
The telescope can observe at 6 bands with wavelengths 20 cm, 12 cm, 6 cm, 3 cm, 1 cm 
and 3 mm. 
Each antenna observes two frequencies simultaneously. 
There are six bandwidths available on Frequency 1 (128, 64, 32, 16, 8 and 4 MHz) and two 
bandwidths on Frequency 2 (128 and 64 MHz). 
The telescope produces $\sim 0.5$ GB of raw data per day, and this is likely to increase 
significantly with future telescope upgrades.

In the rest of this section we outline the existing systems for archiving, processing
and visualising ATCA data. 

\subsection{Data archiving}\label{s_da}
Since the commencement of operation of the ATCA in June 1990, a complete record of all 
data observed from the telescope has been maintained offline at the telescope site, 
mostly recorded on CD. 
In conjunction with this, the ATNF maintained a record of the project proposals for 
observations on the telescope --– the {\it Projects database} --- and a short form of the 
observation parameters for each day’s observing --– the {\it Positions database}. 
After a proprietary period of 18 months, in which the observing team has sole access to 
the data obtained in an observation, the data is made publicly available.
Astronomers can search for observations on the ATNF webpage, submit the details 
of the observation data required via e-mail and have a CD containing the data prepared 
for them at nominal cost.

\subsection{Data processing}
ATCA Data processing (reduction) is generally performed with one of the standard radio data 
reduction packages; most commonly Miriad \citep{sault95}, but also 
AIPS\footnote{Astronomical Image Processing System (AIPS), http://www.cv.nrao.edu/aips.}
and AIPS++\footnote{Astronomical Image Processing System (AIPS++), http://aips2.nrao.edu.}.
After loading, editing and calibrating the data, the resulting product is an intensity map 
referred to as the dirty image.
At this stage a deconvolution algorithm, usually a variant of {\sc clean} 
\citep{hogbom74}, is required to produce the final image.

At each stage in the process there are a range of parameters that can be set to control
the type of processing performed.
Both general parameters (such as calibration strategy, {\sc clean} method and type of 
data editing) as well as fine-grained parameters (such as calibration solution interval, 
number of {\sc clean} iterations and median filter size) need to be modified to obtain 
the best results. 
Hence data processing is typically a highly interactive process.

There is an existing system operating at the telescope {\tt CAONIS} designed for on-the-fly
imaging of ATCA data. 
However the design of this system makes it difficult to port to current Linux systems.

\subsection{Data visualisation}
The final images from the ATCA are usually visualised using tools such as Miriad or 
{\tt kvis} \citep{gooch96}.
These are well established tools that cover many of the visualisation requirements of 
ATCA observers. 
As mentioned in the previous section, the {\tt CAONIS} system which runs at Narrabri also
allows basic visualisation of images.

\section{The Virtual Observatory}
The umbrella organisation for Virtual Observatory work is the International Virtual 
Observatory Alliance (IVOA).
The IVOA was formed in June 2002 with a mission to 
\begin{quotation}{\it 
\noindent facilitate the international coordination 
and collaboration necessary for the development and deployment of the tools, systems and 
organizational structures necessary to enable the international utilization of astronomical 
archives as an integrated and interoperating virtual observatory.}
\end{quotation}

The IVOA is a collaboration between over 15 member countries including Australia.
The focus so far has been developing the standards required for interoperability between
software developed and data produced in all areas of astronomy. 
Another significant aim is to develop the infrastructure required (networks and 
organisations) for the large scale storage and distribution of astronomical data.

The IVOA working groups address a range of issues such as grid and web services,
data modelling and standards for the data access.
There are also four interest groups
\begin{itemize}
\item Applications IG
\item Astronomy Grid IG
\item Data Curation IG
\item Theory IG
\end{itemize}
which focus on the requirements of particular application domains.

The aim of the Australian Virtual Observatory (Aus-VO) is to provide distributed, uniform
interfaces to the data archives of Australia's major observatories and the archives of 
simulation data.
Aus-VO is a collaboration between many Australian institutions, including the Universities of 
Melbourne, Sydney, New South Wales and Queensland, Monash University, 
Swinburne University of Technology, the Australian National University and Mount Stromlo 
Observatory, the Victorian Partnership for Advanced Computing, the ATNF and the AAO.

There are a range of VO projects underway in Australia, including the development of 
data archives and software for HIPASS \citep{meyer04}, RAVE \citep{siebert04}, 
2QZ \citep{croom04} and SUMSS \citep{bock99}.
The initial focus of most of these projects has been to make data from Australian projects 
widely available within the international community, in a VO compliant format. 
In addition there are several projects investigating novel methods for astronomical 
data mining and data analysis, for example \citet{rohde05} apply machine learning 
techniques to catalogue crossmatching.
The Melbourne University group has also been setting up infrastructure such as a registry
for Australian web services and data archives.

\section{The Australia Telescope \\ Online Archive}
In June 2003, a joint project between the ATNF and the CSIRO ICT Centre was commenced to 
make the ATNF archive data available online as the 
Australia Telescope Online Archive (ATOA). 
This was planned as a new data resource for astronomers, as well as the foundation for the 
development of online data processing systems to make the raw data more accessible to 
non-expert users (see Section \ref{s_pipe}).
The construction of the ATOA required the copying of the offline archive (at the time, 
$\sim 2700$ CDs, totalling $\sim 1.7$ TB) from the telescope site to Canberra where 
the online archive was to be developed, creating a meta-database describing the data, and 
making a web front-end to search and download the data.

The database consists of two parts.
The first is the raw data from the telescope (RPFITS files) which is stored as normal files
on the host system. 
In addition there is a relational database which stores all the metadata 
(discussed in Section \ref{s_meta}). 
The `vital statistics' of the ATOA are shown in Table \ref{t_atoa}.
The current rate of growth of the archive is $\sim 0.5$ Gb/day.
However this is likely to increase significantly in the future as new instruments
come online.
To maintain an growing archive (rather than a static one) it is necessary to ensure the 
RPFITS files are stored in a readily accessible way (currently on a RAID system) that is 
easily distributed over a number of drives.
Also, that the database itself is easy to update in a robust manner.
The ATOA was made publicly available in December 2004 and
can be accessed from {\tt http://atoa.atnf.csiro.au}.
\begin{table}[ht]
\begin{center}
\caption{ATOA Statistics}\label{t_atoa}
\begin{tabular}{lr}
\hline 
Projects & 2261 \\
Files & 57147 \\
Sources & 128111 \\
Metadata size & $\sim 4$ Gb \\
RPFITS data size & $\sim 2$ Tb \\
Growth rate & $\sim 0.5$ Gb/day \\
\hline
\end{tabular}
\end{center}
\medskip
\end{table}

\subsection{Metadata}\label{s_meta}
Metadata is simply data which describes other data, for example the project code or the 
name of the primary calibrator.
The meta-database for the ATOA consists of three main parts; the contents of the original 
ATNF online Projects database, metadata describing the observation that is extracted directly 
from the raw data files produced by the telescope’s software, and metadata inferred from all 
of the available data sources to assist in the automation of reducing the telescope’s raw data
to images.
The types of metadata used in the ATOA are summarised in Table \ref{t_meta}.
\begin{table}[ht]
\begin{center}
\caption{Metadata in the ATOA}\label{t_meta}
\begin{tabular}{ll}
\hline Metadata source & Examples \\ \hline
Projects database & proposal; observer name\\
                  & country; institution \\
Positions database$^*$ & source names \& positions \\
                   & observing band; receivers \\
RPFITS files & scans; polarisations \\
             & array configuration \\
Inferred & calibrator names \& roles\\
         & calibrator--target matches \\
\hline
\end{tabular}
\end{center}
$^*$ The positions database is included in our data model, and some of the metadata
is used to reconstruct the observation metadata. However, it is not actually loaded 
into the ATOA.
\medskip\\
\end{table}

Most of the metadata available in the ATNF Positions 
database is also available from the metadata in the raw data files, and is finer-grained, 
since the Positions data is a daily summary, while the file metadata is available for each 
telescope pointing. 
The inferred metadata in the ATOA is `value added' information that is automatically
determined from the existing metadata, for example the calibration role of each source 
(primary calibrator, secondary calibrator, target, etc). 
This is discussed further in Section \ref{s_roles}.

\subsection{A data model for the ATCA}\label{s_dm}
A data model is a comprehensive scheme describing how data is to be represented, for 
manipulation by humans or computer programs.
Data models are critical for planning how data will be organised within a database as they
describe all the relationships between the different entities.

\begin{figure*}[ht]
\begin{center}
\resizebox{16cm}{!}{\rotatebox{90}{\includegraphics{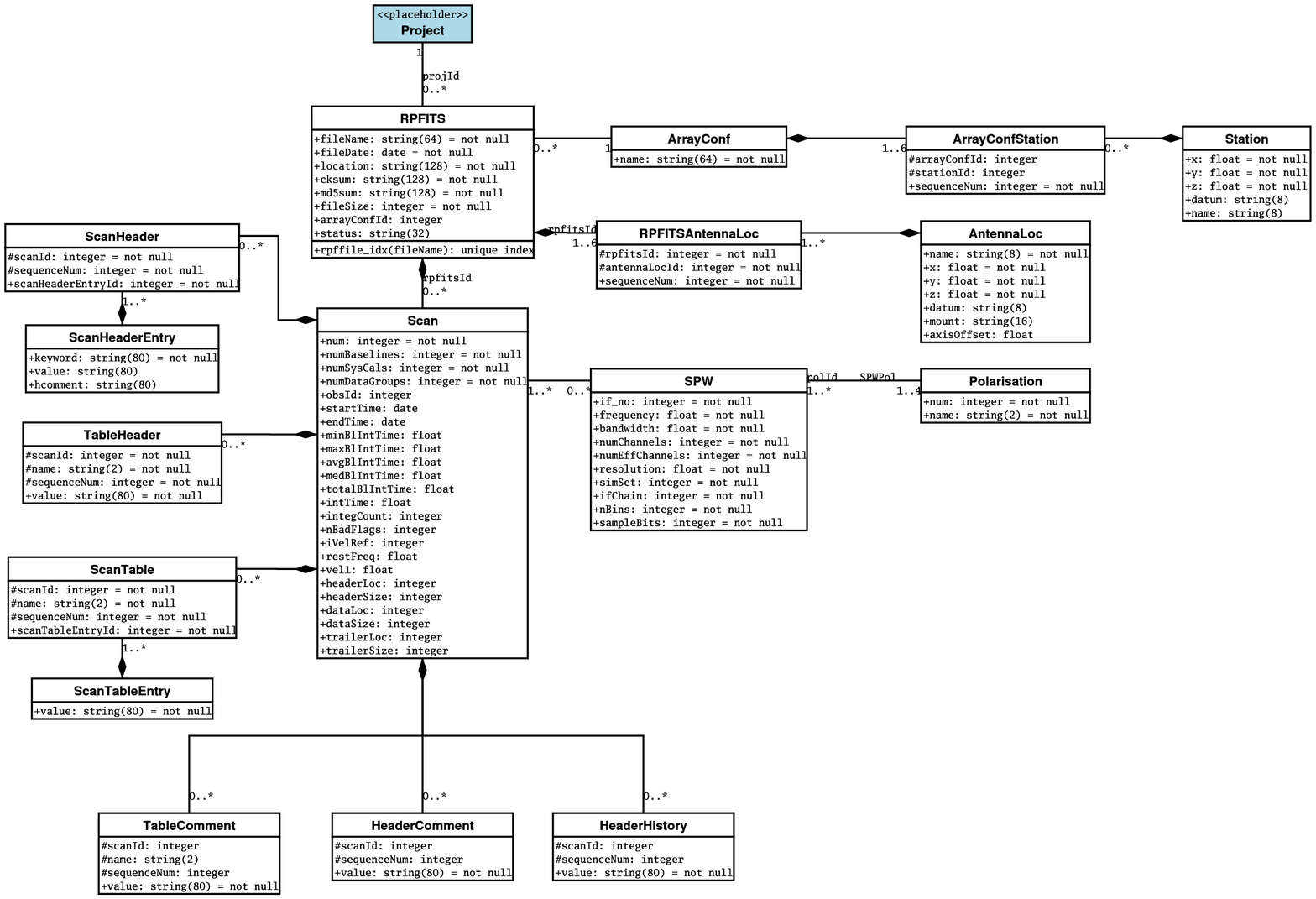}}}
\caption{A portion of the data model for the ATCA. For an explanation of the notation
see Section \ref{s_dm}. The complete data model is available from the ATOA website:
{\tt http://www.atnf.csiro.au/computing/web/atoa/implementation.html}.}\label{f_dm}
\end{center}
\end{figure*}
A section of our data model for the ATCA is shown in Figure \ref{f_dm}.
We now briefly explain the UML (Unified Modelling Language) notation used in the data model.
Each box contains an entity (e.g. {\tt Scan}) that has been identified in
the metadata. 
Each entity has attributes (e.g. {\tt restFreq}), each of which are of a specified
data type (e.g. {\tt float}).
Associated entities are connected to each other with lines, which also specify the
cardinality of the relationship. For example \\
\begin{minipage}{8cm}
\begin{center}
\resizebox{7cm}{!}{\includegraphics{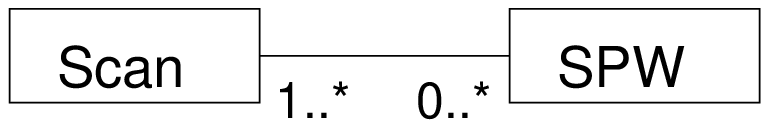}}
\end{center}
\end{minipage}
should be read as {\bf ``A scan has 0 or more spectral windows. A spectral windows has 1 or
more scans.''}

The development of a data model that covers the whole of astronomy is an ongoing project 
within the international VO community. 
We have contributed this data model to the IVOA Data Model WG as an example of a data 
model for radio astronomy.
For more information on this topic, see the IVOA Data Modelling 
website\footnote{{\tt http://www.ivoa.net/twiki/bin/view/IVOA/IvoaDataModel}}.

The ATOA archive database structure is created directly from the
definitions in the ATOA data model. Parts of the data model contain
information for specific database implementations so that all of the
implementation-specific parts of the database creation are handled in
this process.
The data model in Figure \ref{f_dm} corresponds to the part of the data model
that describes the metadata contained directly in the archive RPFITS
files. The data model for the inferred data is available from the ATOA
web pages\footnote{{\tt http://www.atnf.csiro.au/computing/web/atoa/implementation.html}}.

\subsection{Implementation}
The ATOA web interface was implemented as a Java 
(ver. 1.4.2)\footnote{{\tt http://java.sun.com}} 
application and is hosted using the Apache Tomcat 
(ver. 5.0.28)\footnote{{\tt http://tomcat.apache.org}} web container. 
Relational database services are provided by an Oracle 9i instance 
running on the same machine. 
A web based interface was chosen so as to maximise interoperability and provide
easy access to users.
For example, RPFITS files may simply be downloaded by  
constructing a suitable URL for the ATOA file server.
This allows files on the server to be downloaded by a Web browser, by
command-line programs that allow users to fetch the data referred to by
a URL, or by application programs using libraries that allow a URL to be
opened in a similar way to a file on a local file system.

The user interface centres around two main web pages: the query  
page which allows users to specify criteria for selecting RPFITS  
files from the archive, and a results page which provides the means  
for users to inspect the metadata of matching files and download  
particular files if desired. 
The results page initially presents the  
user with a broad, global view of the query results in tabular form  
listing details such as file name, file size, principal investigator  
and array configuration. 
The user may also interactively `drill-down'
for a more detailed view of any file in the list. 
RPFITS files can be downloaded individually or in batches.

As mentioned in Section \ref{s_da} ATCA data has a proprietary period of 18 months,
in which it is only accessible to members of the project team.
Authorised access is supported for data within the proprietary access
period.                                                                                         
If a user wishes to access proprietary data they must first go
through a manually verified authentication process after which a password is issued to the 
Principal Investigator for that project.
In the future we plan to replace this authentication and
authorisation method with a streamlined system that links the new ATNF
proposal system, OPAL\footnote{{\tt http://opal.atnf.csiro.au/}}, and its authentication 
database to the ATOA.
Users will then be given access to proprietary data by using their
OPAL credentials based on the projects they are associated with in
the OPAL system

The ATOA web server and database are hosted on a Dell PowerEdge 750
running Debian Linux 3.0. The host has a 2.8GHz Pentium 4 processor,
2GB of RAM and is attached to a 3 terabyte Apple Xserve RAID for
archive storage.

\section{A data processing pipeline framework}\label{s_pipe}
The data products in radio astronomy are often less accessible to the non-expert than
those in other domains such as optical astronomy.
It requires a reasonably high level of domain expertise to process the raw data and 
produce an image. 
Obviously for carrying out detailed scientific analysis it would be necessary to 
develop this expertise, or collaborate with a radio astronomer. 
However in an era of multiwavelength astronomy, astronomers expect to download and 
compare data from a variety of telescopes, at a variety of wavelengths. 

With this in mind we have developed an automatic pipeline for people who want to 
quickly inspect the data in the ATOA, to see if it was suitable for further processing.
One of the aims of this project was to test the viability of `driving' the pipeline
using the metadata discussed in Section \ref{s_meta}. In other words the pipeline should
make decisions about what kind of processing to do --- both on a general 
(e.g. continuum/spectral line) and specific level (e.g. number of {\sc clean} iterations).

In this section we discuss the development of extra metadata required to driving the pipeline,
in particular the calibration process.
We then outline our prototype pipeline which can process
single pointing continuum data from the ATOA and is available for testing at 
{\tt http://atoa.atnf.csiro.au/test}.

\subsection{Metadata for automatic \\ processing}\label{s_pmeta}
The metadata in the Project and Position databases, while providing information about 
which astronomical sources have been recorded in an observing session, does not (in general) 
provide any information about the role that the observer intended the source to play in the 
observation (eg. primary calibrator, target source).
This would be relatively easy to record in a new system, but as we are dealing with 
existing data we had to infer the roles of sources.

Another problem for automatic processing is the grouping of data into valid `observations'.
An expert would typically choose an appropriate subset of files from the archive to image.
However, a non-radio astronomer may choose an subset that contains files that should be imaged
separately, or files that contain data that should be ignored entirely.
Although it is impossible to deal with all cases, our aim was to have the pipeline 
group together the selected data in such a way that an image could be made in at 
least $80\%$ of cases.
A wide range of observation types can be recognised and characterised using the meta-database
but are not yet processed by the prototype pipeline (e.g. millimetre and spectral line 
observations).

In the following section we discuss how we assign the source roles within an observation, and
the algorithm we used to match target sources with the appropriate calibrators.

\subsection{Determining source roles}\label{s_roles}
While matching target sources with their calibrators would be straightforward for an
astronomer it is a challenge for an automatic system.
In a typical (simple) observing session the primary calibrator is recorded for a short period 
at the start or end of the observing session; and alternating pointings are made to the 
secondary, and the source of interest, or target.
However there is a great variety of different ways that the observer can choose to structure
their observations.
If an observation contains more than one target, 
the targets may share, or have distinct, secondary calibrators, depending on their 
separation in the sky. 
There may be several secondary calibrators for each target, and the same source may be 
used for primary and secondary calibration. 
In addition, some observers use secondary calibrators that are not in the list of recommended
calibrators, and that list has itself changed over time.

In order to classify the sources in an observing session the following metadata is used
\begin{itemize}
\item The locations and names of sources extracted from the raw telescope data
\item The times and durations of the source pointings
\item The names and locations of the four primary calibrators commonly used at the ATCA
\item A recent ATCA catalogue of recommended secondary calibrators
\item Names of sources extracted from project titles
\item A pre-assembled list of possible calibrator sources
\end{itemize}

Once the source roles have been determined, the proximity in the sky and the proximity 
in observation time of the targets and their secondary calibrators are used to match targets 
with their respective calibrator(s). For each target pointing, a weight is calculated for 
each secondary calibration pointing made within two hours of observation of the target pointing:
\begin{displaymath}
w_{t,s} = \Sigma_{P} \Sigma_{S} e^\frac{-(3a/a_{max})^2}{2} 
e^\frac{-(3\Delta t / \Delta t_{max})^2}{2} 
\end{displaymath}
\begin{displaymath}
\Delta t_{t,s} < \Delta t_{max}
\end{displaymath} 
where $S$ is the set of candidate secondary calibrators, $a$ is the angular separation 
between the target and the secondary calibrator, $a_{max}$ is the maximum desirable 
separation between the target and the secondary calibrator (and is a function of the 
observing frequency band). 
$\Delta t$ is the separation of the time midpoints of the target and calibrator 
pointings and $\Delta t_{max}$ is the maximum desirable time separation (two hours).
The summation is over all pointings at a target ($\Sigma_P$) and all secondary 
calibrators within two hours of a target pointing ($\Sigma_S$).
The $w_{t,s}$ are used to select suitable secondary calibrators for the respective 
targets from the calibrators whose $w_{t,s}$ weights dominate for a particular target.

This procedure constructs the metadata required for continuum imaging at centimetre 
wavelengths.
The algorithm works well in general, but there are some problematic cases, for example 
where the target is a source from the secondary calibrator catalogue.

\subsection{Implementation}
The underlying processing of the ATCA data is carried 
out using the Glish scripting language in AIPS++.
The ATOA imaging Web Services interface was constructed using the Apache Axis tools 
(ver. 1.1)\footnote{{\tt http://ws.apache.org/axis}}, and 
interfaces to the processing scripts through a Perl 
(ver. 5.4.8)\footnote{{\tt http://www.perl.com}} script that deals with the control of 
the execution of the Glish scripts.

The pipeline client is written using Python (ver. 2.3)\footnote{{\tt http://python.org}}, 
and the SOAPpy web services tools 
(ver. 0.11.3)\footnote{{\tt http://pywebsvcs.sourceforge.net}}. 
There were some minor, but difficult to find, problems in interoperation between the SOAPpy 
tools and Apache Axis; the data structures used in the web services calls are possibly more 
complicated than had been previously used between the two web services implementations. 
Documentation in both was not as informative or complete as it might have been.

The pipeline web services can be configured to run directly on the server host, or be 
directed to run on other machines through a batch queuing system, since some stages 
in the pipeline can run for several CPU minutes. 
We used the OpenPBS Batch Queuing System (ver. 2.3)\footnote{{\tt http://www.openpbs.org}} 
for queue management, but unfortunately it has no mechanism for 
reporting job completion to another program. 
After processing for a web service completes, the batch job doing the
processing sends a completion message to the program invoked by the web
service that controls the execution of the processing for the service.
However, at this point, the batch processing system has not yet
transferred the job's output data back to the pipeline server. The
control program then polls the PBS batch queue at five second intervals
to ensure that the batch job has completed.


The raw data from the ATOA, all the intermediate files from the data processing, the log files, 
and the resulting images are stored temporarily on the pipeline server. 
The first web service call made by a pipeline client reserves a private location for storage, 
and requests a lifetime for the storage. The pipeline server has a configurable
maximum lifetime, and the stored data will be deleted after this time expires.
Only clients who have the name of the storage area (a randomly generated string) can
access it.
There is no quota on the storage use of any individual temporary storage
area. 
However, a quota may be imposed on the total amount of storage available to
all active storage areas.

\begin{figure}[ht]
\begin{center}
\resizebox{7cm}{!}{\includegraphics{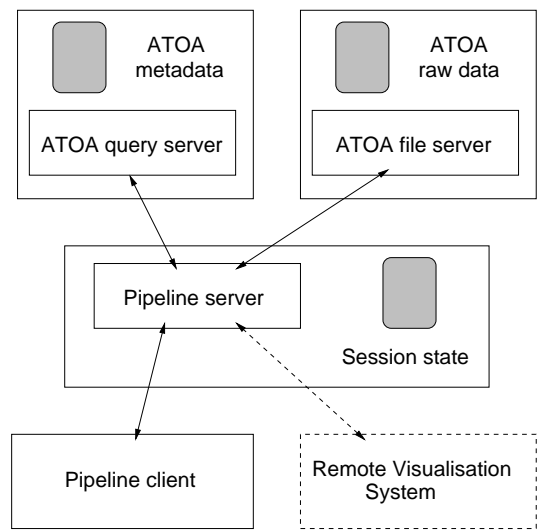}}
\caption{System architecture. This schematic shows the relationship between the three
tools we have developed.}\label{f_arch}
\end{center}
\end{figure}
The ATOA and pipeline web services 
return a URL for the generated images to the end-user's system. 
This allows the URL to be passed on to the Remote Visualisation System (see Section 
\ref{s_rvs}) image viewing system so that
the image can be viewed online while it is still resident on the
pipeline server. 
Figure \ref{f_arch} show the overall system architecture, in particular how the ATOA, 
pipeline and RVS interact.

\section{The Remote Visualisation \\ System}\label{s_rvs}
\begin{figure}[ht]
\begin{center}
\resizebox{7cm}{!}{\includegraphics{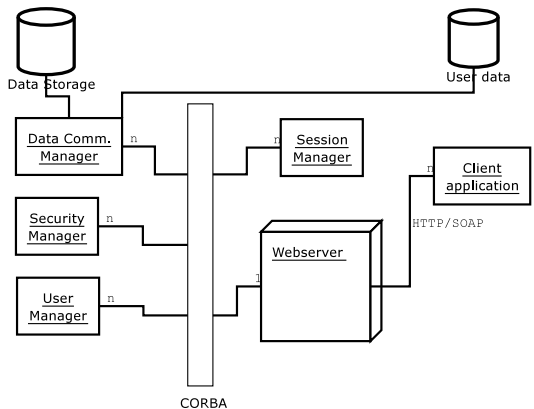}}
\caption{RVS system architecture.}\label{rvs_arch}
\end{center}
\end{figure}

The Remote Visualisation System (RVS) was designed to enable visualisation of 
and interaction with large astronomical images in the context of the VO. 
As opposed to other VO image displays, such as CDS Aladin \citep{fernique04}, 
RVS does not require the user to download the data to the client machine. 
Furthermore it provides rendering of image cubes, such as spectral-line cubes 
created from ATCA data.
The RVS server accepts FITS images - which can be compressed - through local 
{\tt file://} URLs and remote http or ftp access. 
The data should be co-located with or at least be available to the server on 
high bandwidth connection, while it places no such requirements on the client. 
Only minimal data transfer to the client is necessary and this is independent 
of the size of the source data set. 
The server-side architecture is distributed to enable workload sharing and 
extensibility. 
RVS makes use of several software components: CORBA to make it distributed, 
AIPS++ as the image rendering component and Java for the web services
and client.
The architecture of the RVS system is shown in Figure \ref{rvs_arch}.

RVS is exposed through a web service interface using the standard Web Service
Description Language (WSDL ver. 1.1)\footnote{{\tt http://www.w3.org/TR/wsdl}}.
This can easily be integrated into custom applications.
Several client applications make use of the web service interface; the 
RVSViewer - a traditional image viewer, a thumbnail service - providing 
preview images and a session viewer. 
The session viewer connects to an existing RVSViewer via a key. 
Multiple instances can be run at the same time, making it a possible to use 
it as a conferencing tool where people can observe and interact with the data.
The ATOA pipeline re-uses the existing RVSViewer client by passing it the file 
location of the output image.

RVS is not specific to the ATOA or prototype pipeline and there are plans to
use it for all ATCA archives.
It has been successfully tested on images and data cubes from various surveys and has
good performance on large datasets. 
For example a 1.5 Gb data cube from the Galactic All-Sky Survey (GASS) \citep{mcclure05} 
takes about one minute to load.
Compare this with downloading the full cube from say the U.S. to Australia which could 
take $\sim 1-2$ hours.
For more information and direct access to RVS, see {\tt http://www.atnf.csiro.au/vo/rvs/}.

\section{Discussion}
The ATOA has been public since December 2004.
We hope that it will encourage the reuse of ATCA data for projects other than 
those it was originally intended for. 
The framework used for the ATOA could easily be extended to include data from other 
telescopes and can be updated as additional metadata is required.

The most significant improvement of the ATOA over existing online archives (such 
as NRAO\footnote{{\tt http://archive.nrao.edu/archive/e2earchive.jsp}} and 
MAST\footnote{{\tt http://archive.stsci.edu/}}) is the data delivery mechanism.
Most existing archives do not support on demand delivery
of data over the web, instead requiring the user to submit a form
requesting files that then have to be transfered to a publicly
accessible ftp site or to other media (such as CD) for physical
delivery. 
In the ATOA, the batch downloading of multiple files is handled by a 
streaming TAR or ZIP archiving algorithm that performs dynamic archiving as 
files are streamed over the web, requiring no additional disk space on the
server for these operations.

In developing the ATCA data model and considering the type of metadata required for automatic 
processing we identified several new metadata types that would be useful to store in the 
RPFITS files. 
As a result the following fields have been added to the RPFITS files and will be available in 
all future ATCA data:
\begin{itemize}
\item four calibrator codes
  \begin{description}
  \item[C] (standard phase calibrator)
  \item[F] (primary flux calibrator)
  \item[B] (bandpass calibrator)
  \item[P] (pointing calibrator)
  \end{description}
\item Pointing offsets
\item Weather data: added rain gauge and phase rms and difference
\item Attenuator settings at start of scan
\item Subreflector position
\item Correlator configuration
\item Scan type
\item Coordinate type
\item Line mode
\item CACAL counter 
\end{itemize}
These will help both automatic processing systems and astronomers assess the data quality
in the observations they are interested in.
A full e-logbook system will be used in the future as currently the logs are all stored on
paper at the telescope and hence are not easily accessible to ATOA users.

We have developed a prototype pipeline for processing of raw data for single-pointing 
continuum images.
This is attached to the ATOA to provide an improved service for users of the ATOA. 
At this stage the image quality is suitable for previewing the data in archive to see
if it is of interest.
Further manual processing would then be required to obtain images of scientific quality.

A significant challenge in developing the ATOA and the prototype pipeline were integrating
pre-existing software with modern software tools.
For example, the Glish scripting language has no web service libraries and so an extra
layer had to be developed between the data processing level and the web services.
If re-implementing from scratch, a language such as Python would be a better alternative 
for developing the pipeline.

In developing these tools we have started to explore the techniques necessary for 
astronomical software development in the VO era. 
This is essential for future telescopes and surveys that Australia will produce.
Making access to existing Australian data as easy as possible will maximise
its use in the international community.

\section*{Acknowledgments}
The authors would like to acknowledge the software development done
on the RVS project, primarily by Anil Chandra and also by Praveena Tokachichu.
The ATNF side of the prototype pipeline and ATOA development was managed by
Neil Killeen and Jessica Chapman.
Vince McIntyre contributed extensively to all three projects, in particular
in setting up the hardware required.

A number of ATNF staff put in significant effort to get the ATOA set up, 
in particular Robin Wark, Bob Sault and Mark Wieringa.
Warwick Wilson and Mark Wieringa implemented the changes to add extra 
metadata to the RPFITS files.

From the ICT Centre, Robert Power made the initial data model designs,
the ATOA data loader software and ATOA query front end.  Geoff Squire and
Bella Robinson made significant contributions to the prototype pipeline.


\end{document}